\documentclass[conference]{IEEEtran}
\IEEEoverridecommandlockouts
\usepackage{cite}
\usepackage{amsmath,amssymb,amsfonts}
\usepackage{algorithmic}
\usepackage{graphicx}
\usepackage{textcomp}
\usepackage{xcolor}
\def\BibTeX{{\rm B\kern-.05em{\sc i\kern-.025em b}\kern-.08em
    T\kern-.1667em\lower.7ex\hbox{E}\kern-.125emX}}

\usepackage{pgfplots}
\pgfplotsset{compat=1.17}
\usepackage{pgfplotstable}
\usepackage{booktabs}

\pgfplotstableset{
  every head row/.style={before row=\toprule,after row=\midrule},
  every last row/.style={after row=\bottomrule},
  fixed,precision=2,
}

\begin{document}

%
\title{Towards an Efficient Simulation Approach for Transmission Systems with ICT Infrastructures
\thanks{This work has been prepared with the support of the Belgian Energy Transition Fund, project CYPRESS (https://cypress-project.be).}}

\author{
\IEEEauthorblockN{Frédéric Sabot, Pierre-Etienne Labeau, Jean-Michel Dricot, Pierre Henneaux}
\IEEEauthorblockA{Université libre de Bruxelles \\
Brussels, Belgium\\
\{frederic.sabot, pierre.etienne.labeau, jean-michel.dricot, pierre.henneaux\}@ulb.be}
}


\section*{Erratum}

It has been found that a bug in our co-simulation implementation was causing unnecessary I/O operations which significantly slowed down the co-simulations. The table below is the updated version of Table~\ref{tab:computationTime} obtained after fixing the bug.

\begin{table}[h]
\centering
\begin{tabular}{@{}lll@{}}
\toprule
Time precision & Self-consistent simulation & Co-simulation \\ \midrule
10 ms     & 4 s                        & 7 s          \\
1 ms      & 7 s                        & 16 s         \\ \bottomrule
\end{tabular}
\end{table}

This updated table shows a significantly smaller gap between the computation times of ``classical" co-simulations and of our proposed ``self-consistent" simulations. This smaller gap might thus not justify the additional complexities and assumptions introduced by our method.

\cleardoublepage 

\maketitle

\begin{abstract}
With the transition towards a smart grid, Information and Communications Technology (ICT) infrastructures play a growing role in the operation of transmission systems. Cyber-physical systems are usually studied using co-simulation. The latter allows to leverage existing simulators and models for both power systems and communication networks. A major drawback of co-simulation is however the computation time. Indeed, simulators have to be frequently paused in order to stay synchronised and to exchange information. This is especially true for large systems for which lots of interactions between the two domains occur. We thus propose a self-consistent simulation approach as an alternative to co-simulation. We compare the two approaches on the IEEE 39-bus test system equipped with an all-PMU state estimator. We show that our approach can reach the same accuracy as co-simulation, while using drastically less computer resources.

\end{abstract}

\begin{IEEEkeywords}
Co-simulation, cyber-physical systems, self-consistent simulation
\end{IEEEkeywords}


\section{Introduction}
\label{sec:Intro}



The energy transition is expected to lead to smarter electric power systems taking the form of Cyber-Physical Systems (CPS) in which electrical power grids are strongly interlinked with a growing number of information and communication systems. This transition brings many opportunities but also new cyber-physical threats (cyber-attacks, ICT infrastructure unreliability, etc.) whose associated risk should be assessed and minimised. One way of assessing the impact of cyber-physical threats is to combine power system and communication network simulators to perform so-called co-simulations. The latter allows to reuse existing simulators from both domains and to accurately model the interaction between the two domains.


Challenges of co-simulation are however synchronisation of the simulators, data exchange between the simulators, and computation time~\cite{CoSimChallenges}. Significant work has been performed to overcome these challenges. These efforts led to a number of co-simulation frameworks upon which (almost) any simulator can be synchronised. 
Several ad hoc approaches were also developed to synchronise specific simulators. We refer to~\cite{VogtCosimReview, GridAttackSim, ToolsForCosim} for a complete review of the tools available for the co-simulation of power systems with other systems (including ICT systems).

Despite this work, computation speed remains a challenge. 
This is because simulators frequently have to be synchronised, i.e. paused to exchange information and then restarted. Indeed, when an event occurs in a simulator, this information is not directly available in the other simulator(s). This causes causality or synchronisation errors. It is thus necessary to frequently synchronise the simulators to mitigate those errors. This leads to high computation times. For example, Ref.~\cite{GECOcomputationTime} shows that when reducing the time step of the power system simulator from 10 to 1~ms, additional discrete power system events occur and have to be propagated to the communication simulator. This leads to an increase of computation time by a factor 20. Note that, in this case, the simulators are only synchronised when an event occurs, not at each time step of the power system simulator.

In order to avoid this large computational burden, we propose self-consistent simulation as an alternative to co-simulation. In this approach, the simulators are run sequentially instead of simultaneously which removes the need for synchronisation. To the best of the authors' knowledge, this is the first time an alternative simulation approach for CPS is proposed.





Section~\ref{sec:methodo} explains our proposed methodology. Sections~\ref{sec:CaseStudy} and~\ref{sec:Cosim} respectively present our test case and the co-simulation platform in which it is implemented. Section~\ref{sec:Results} compares the results obtained with self-consistent and co-simulation, with section~\ref{sec:Results-Monitoring} focusing on monitoring, and section~\ref{sec:Results-Control} on control applications. Finally, section~\ref{sec:Conclusion} concludes with a summary and perspectives.

\section{Methodology}
\label{sec:methodo}

The basic idea (but not a required hypothesis) underlying our methodology is the fact that the traffic generated by most smart grid applications (Supervisory Control And Data Acquisition (SCADA), Phasor Measurement Units (PMUs), Advanced Metering Infrastructure (AMI), etc.) is independent of the state of the electrical grid. A notable exception is teleprotection that is by nature based on the crossing of thresholds by electrical variables. However the associated traffic volume is often negligible compared to the traffic of other smart grid applications~\cite{TrafficVolumeBook}.

\begin{figure}
    \centering
    \includegraphics[scale=0.605]{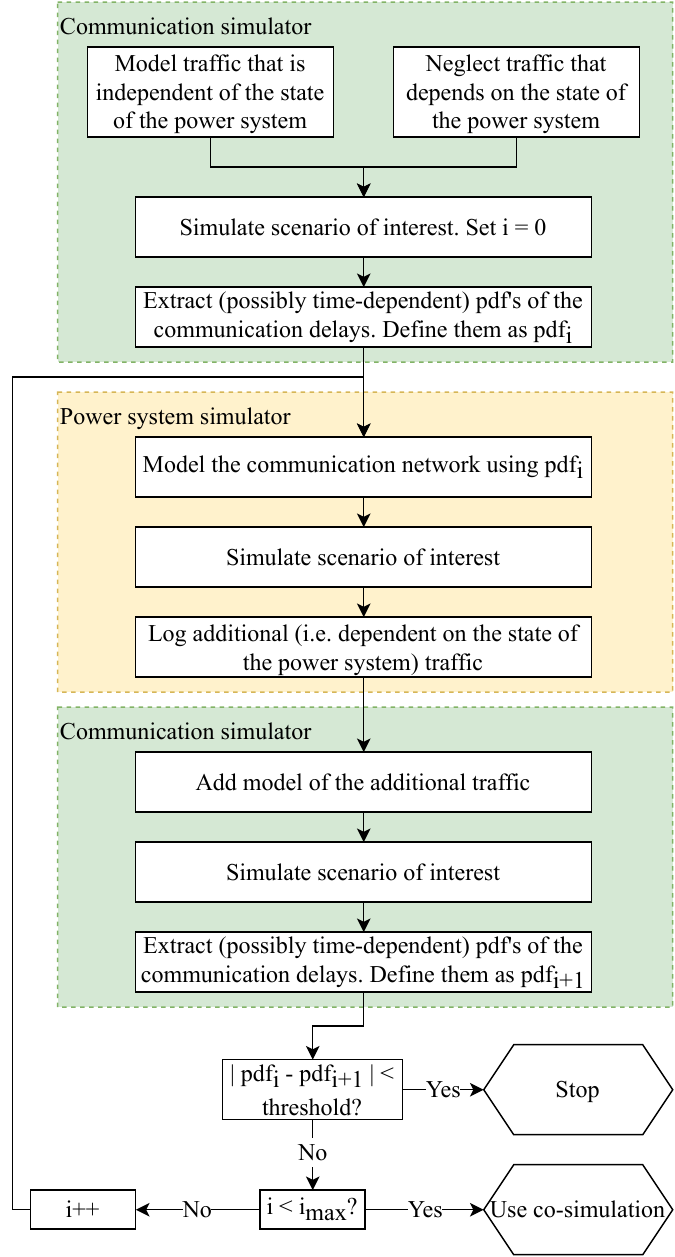} 
    \caption{Flowchart of the proposed methodology}
    \label{fig:flowchart}
\end{figure}

The proposed method thus simulates CPS by first neglecting the impact of the power system dynamics on the communication network. 
The methodology consists of two parts and is illustrated by a flowchart in Fig.~\ref{fig:flowchart}. In the first part (corresponding to the first two blocks in the flowchart), the CPS is simulated using the communication network and power system simulators separately. In the second part (corresponding to the last block in the flowchart), the validity of this simulation is checked.

In the first part, we simulate CPS by (i) simulating the communication network neglecting the traffic that depends on the state of the power system, (ii) retrieving the probability density function (pdf) of the delays between nodes of interest as well as the associated probability of packet loss, (iii) injecting the pdf's of those delays in the power system models, and (iv) simulating the power system.

Then, in the second part, we verify the accuracy of this CPS simulation by (i) logging the ``additional traffic'', i.e. the traffic that depends on the dynamic evolution of the power system, (ii) injecting this additional traffic in the communication network simulator, and (iii) checking if this additional traffic significantly impacts the performance of the communication network. If it is not significantly impacted, then our simulation is valid. Otherwise, it is not, and we might need to use co-simulation instead.

As an indicator of the performance of the communication network, we use the pdf's mentioned in point (ii) of the first part of the methodology. We thus recompute these pdf's from the communication network simulation that includes the additional traffic, and compare them with the ones obtained without the additional traffic. If the norm of the difference between the two set of pdf's is lower than a threshold, we deduce that the performance of the communication network was not significantly affected by the additional traffic. An example of norm is the maximum over all times and communication paths of the average delay difference.



As such, our methodology would still require the use of co-simulation for the cases where the additional traffic has a non-negligible impact on the performance of the communication network. To increase the number of cases for which co-simulation can be avoided, we use a self-consistent iteration scheme. In other words, for the cases where the pdf's used as a model of the communication network in the power system simulator (i.e. the pdf's initially computed without the additional traffic) differ from the pdf's computed with the additional traffic, we reinject the second set of pdf's in the power system simulator. The power system simulator then computes a new estimation of the additional traffic that leads to a new estimation of the pdf's that can be reinjected in the power system simulator. This iterative process stops when the pdf's used as input of the power system simulator do not significantly differ from the pdf's obtained at the output of the CPS simulation.

In order for our methodology to be more computationally efficient than co-simulation, the number of iterations should be limited. However, as explained in section~\ref{sec:Cosim} and illustrated by Table~\ref{tab:computationTime}, the computation cost of co-simulation can be one order of magnitude higher than the sum of the computation cost of the individual simulators. This gives some margin on the number of iterations our method can use in order to stay competitive with co-simulation.

\section{Case study}
\label{sec:CaseStudy}



\begin{figure}
    \centering
    \includegraphics[width=0.822\linewidth]{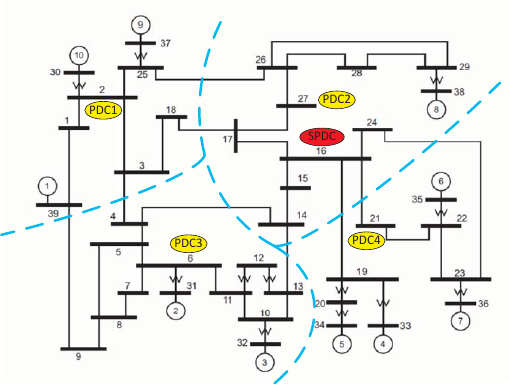}
    \caption{IEEE 39-bus test system with an all-PMU state estimator~\cite{GECOtestcase}}
    \label{fig:IEEE39}
\end{figure}

The test case used in this work is based on~\cite{GECOtestcase} and is shown in Fig.~\ref{fig:IEEE39}. It consists of the IEEE 39-bus test system~\cite{IEEE39} and an all-PMU state estimator. Dynamic data has been taken from~\cite{IEEE39transient}. Loads are considered as constant impedance loads.

All buses are equipped with a PMU. These PMUs measure complex voltages with a sample rate of 30 times per second, time-tag those measures, and send them to the nearest Phasor Data Concentrator (PDC). PDCs are located at buses 2, 6, 21 and 27. When a PDC gathers all measurements for a given timestamp, it sends them to a Super PDC (SPDC). A PDC will also send measurements for a given timestamp if it has to wait longer than a predefined time to receive all measurements (a measurement that arrive after the wait time is disregarded). We use the relative wait time logic in this work, i.e. the wait time counter of a PDC is triggered upon receiving the first measurement for a given timestamp~\cite{Monolithic}. The SPDC uses the same logic. The SPDC and PDCs have a maximum wait time of 100~ms.

Regarding the communication infrastructure, a router is placed at each bus, and a full-duplex link with a bandwidth of 800~kbps is installed in parallel to each transmission line. The propagation delay in links is computed assuming a refractive index of 1.5. The line lengths are deduced from their impedance assuming lines have a linear impedance of 0.3~\(\Omega\)/km (typical of 345~kV lines), as in~\cite{LineLength}. All packets have a size of 500 bytes. The UDP communication protocol is used as proposed in the IEEE C37.118.2 standard~\cite{StandardC37-118-2}.

\section{Co-simulation}
\label{sec:Cosim}

In this work, we use the open-source transmission system simulator Dynawo~\cite{dynawo}, communication network simulator ns-3~\cite{ns3}, and co-simulation framework HELICS~\cite{helics}. The same simulators are used to implement the self-consistent simulation approach. The implementation of the co-simulator and test case is available on https://fsabot.ulb.be/.

Power system and communication network simulators use different time management approaches. Indeed, power systems are modelled using differential-algebraic systems of equations and consider time as a continuous variable. On the other hand, communication network simulators model the evolution of a system as a sequence of events. Events occur at discrete points in time and no modification of the system state occur in between events.

An important decision that one has to take when implementing a co-simulator is thus the choice of the synchronisation method. There are two main types of synchronisation methods: time stepped and event-based~\cite{VogtCosimReview}. In the first method, simulators are synchronised (i.e. are paused to exchange data, and then restarted) at predefined (usually periodic) time steps. In the second method, simulators are synchronised each time an event occurs in one of the simulators. In the communication network simulator, an example of event is a packet arriving at a router or at its destination. In the power system simulator, events are usually threshold crossings (e.g. the current in a line crossing its maximum acceptable value), or in our case, the periodic sampling of PMUs.

Since our goal is to use the co-simulation results as a reference to validate our self-consistent simulations results, the event-based synchronisation approach was preferred. Indeed, time stepped synchronisation causes so-called synchronisation errors. When an event occurs in a simulator, the other simulator(s) will not be aware of this occurrence until the next synchronisation step. Events are thus perceived as ``delayed'' by the other simulator(s).

However, a drawback of event-based synchronisation is that the number of synchronisation steps can become very large. In our test case, we have 39 PMUs that send messages to the communication network 30 times per second. If we take as a rough estimate that there are in average 3 hops between a PMU and the SPDC, this results in 3500 events and thus 3500 synchronisation steps per second, or in average, a synchronisation step every 0.3~ms. Synchronisation steps would be even more frequent for larger systems. In order to reduce the number of synchronisation steps, we only allow the communication network simulator to create a synchronisation step 1~ms after the current time. No restriction is put for the power system simulator since it does not create as many events\footnote{On the power system side, PMUs are assumed to be perfectly synchronised (i.e. all messages are sent at the same time), so the power system only generates 30 synchronisation steps per second. On the communication side, different propagation delays in the different links lead to messages arriving at different times even if sent simultaneously.}.

\section{Results}
\label{sec:Results}

This section compares results obtained with co-simulation and with our proposed methodology. Section~\ref{sec:Results-Monitoring} focus on monitoring applications while section~\ref{sec:Results-Control} focuses on control applications, i.e. considers that the SPDC will send control commands based on the power system state estimation.

\subsection{Monitoring}
\label{sec:Results-Monitoring}

For monitoring applications, the volume of traffic sent to the communication network is very often independent of the state of the power system. In our considered test case, PMUs send data periodically, and in packets of constant size. It is thus straightforward to model this traffic (but not the values of the measurements) in the communication network simulator without any information from the power system simulator.

\begin{figure}
\centering
\begin{tikzpicture} [scale=0.7]
\pgfplotsset{width=\linewidth,
        legend style={font=\footnotesize}}
\begin{axis}[
    xlabel={Timestamp [s]},
    ylabel= {Delay [ms]},
    legend cell align=left,
    legend style={at={(0.02,0.98)},anchor=north west,
    },
    y filter/.code={\pgfmathmultiply{#1}{1000}}
   ]

  \addplot[only marks, blue, mark=+, legend image post style={mark=*}] table [x=Time,y=2] {Monitoring.txt};
  \addplot[only marks, red, mark=+, legend image post style={mark=*}] table [x=Time,y=6] {Monitoring.txt};
  \addplot[only marks, green, mark=+, legend image post style={mark=*}] table [x=Time,y=21] {Monitoring.txt};
  \addplot[only marks, mark=+, legend image post style={mark=*}] table [x=Time,y=27] {Monitoring.txt};
  
  \addplot[only marks, blue, mark=x, legend image post style={black, mark=+}] table [x=Time,y=2Cosim] {Monitoring.txt};
  \addplot[only marks, red, mark=x, legend image post style={black, mark=x}] table [x=Time,y=6Cosim] {Monitoring.txt};
  \addplot[only marks, green, mark=x] table [x=Time,y=21Cosim] {Monitoring.txt};
  \addplot[only marks, mark=x] table [x=Time,y=27Cosim] {Monitoring.txt};
  \legend{Bus 2, Bus 6, Bus 21, Bus 27, Self-consistent, Co-simulation}

\end{axis}
\end{tikzpicture}
\caption{Comparison of results obtained with self-consistent simulation and co-simulation for monitoring applications. Note that the results of the two approaches are superimposed}
\label{fig:monitoring}
\end{figure}
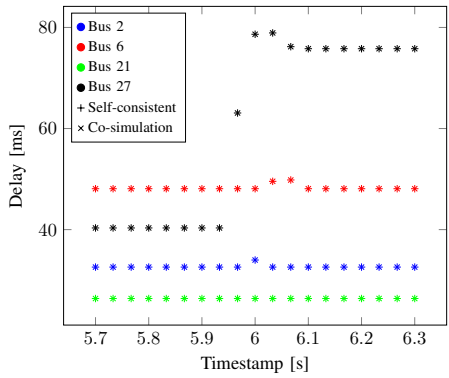

In order to validate our self-consistent and co-simulation implementations, we verify that they give consistent results. To do so, we simulate a failure of the (communication) link between buses 16 and 17 at time \(t\) = 6~s, and we compare the communication delays obtained in the two simulations. Fig.~\ref{fig:monitoring} shows the delays between PMUs sending measurement and the arrival of those measurements at the SPDC for the PMUs at buses 2, 6, 21 and 27 (where the PDCs are located). As can be seen in the figure, the two simulations give almost identical delays.

It should be noted that to ease the comparison between the two simulations, we did not model random phenomena (e.g. errors in links). The delays computed are thus deterministic and do not require pdf's in order to be modelled in the power system simulator. Note also that since there is no additional traffic, our method converges in a single iteration.

\subsection{Control}
\label{sec:Results-Control}

For control applications, traffic might depend on the state of the power system, and thus cannot be modelled entirely without information from the power system simulator\footnote{For some cases, control traffic might be constant. A typical example is inter-area oscillations damping. For these cases, control traffic can be modelled similarly to monitoring traffic.}. We still consider the case study described in section~\ref{sec:CaseStudy}, but we now consider that the SPDC can send control messages.

The scenario used in this study is the loss of generator 3 causing a loss of 650 out of 6140~MW of production. In this scenario, if no action is taken by the SPDC, the frequency will drop below 49~Hz after 2.4~s. This would lead to the activation of under-frequency load shedding relays and thus lead to energy not supplied.




In order to avoid load shedding, the SPDC will send load reduction messages to contracted interruptible services when it detects a fast drop of the frequency. Note that it is not really relevant to study this scenario using co-simulation. Indeed, the frequency only drops below 49~Hz 2.4~s after the generator disconnection. Communication delays, that are usually in the order of dozens of milliseconds, can thus be neglected in this case. So, in order to emulate a more time-critical application, we design the SPDC to operate very quickly. The SPDC is designed to send a load reduction signals to every loads when the average frequency of the received measurements drops below 49.96, 49.92 and 49.88~Hz. The SPDC will thus send a quick succession of control messages as it would in a more time-critical application. At each step, all loads reduce their active and reactive power demand by 2\%.


Now, to implement our self-consistent approach, we need to ``retrieve the pdf's of the delays between nodes of interest'' from the communication network simulator results. In this case, the delays of interest are the difference between the time at which the PMUs measure an average frequency lower than a threshold, and the times at which the loads receives the associated load reduction signal. However, in normal operation (i.e. when no ``additional traffic'' is modelled), there is no traffic from the SPDC to the loads. The trick is thus to consider that for an arbitrary timestamp the SPDC will receive an average frequency lower than the first frequency threshold, and thus send messages to the loads. The time differences between this timestamp and the times at which the loads receive a signal are then used as \(pdf_0\). As previously mentioned, we did not consider random phenomena such as link errors, the pdf's are thus actually Dirac functions. In the rest of this section, we will thus compare the pdf's using only their average. Applying our methodology to cases with uncertainties will be done in future work.

\begin{figure}
\centering
\begin{tikzpicture} [scale=0.7]
\begin{axis}[
    ybar,
    enlargelimits=0.2,
    legend style={at={(0.5,-0.15)},
      anchor=north,legend columns=-1},
    ylabel={Delays [ms]},
    symbolic x coords={Load 8, Load 23, Load 29, Load 39},
    xtick=data,
    ybar=2pt,
	bar width=8pt,
    ]
\addplot coordinates {(Load 8, 47.7452) (Load 23, 33.9576) (Load 29, 56.4154) (Load 39, 64.5101)};
\addplot coordinates {(Load 8, 47.7451) (Load 23, 33.9575) (Load 29, 56.4151) (Load 39, 64.51)};
\addplot coordinates {(Load 8, 47.7453) (Load 23, 33.9578) (Load 29, 56.405) (Load 39, 64.5103)};
\addplot coordinates {(Load 8, 47.7457) (Load 23, 33.9581) (Load 29, 56.4051) (Load 39, 64.5106)};
\legend{\(pdf_0\), \(pdf_1 (1.1)\), \(pdf_1 (1.233)\), \(pdf_1 (1.366)\)}
\end{axis}
\end{tikzpicture}
\caption{Comparison of the averages of the input and output pdf's at the first iteration of the self-consistent simulation. 1.6~Mbps case}
\label{fig:it1-1.6Mbps}
\end{figure}
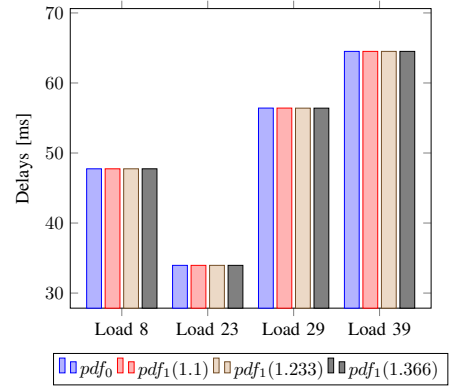

Once \(pdf_0\) is obtained, it can be used to model the communication delays in the power system simulator. The power system simulator is then run. It computes (from the evolution of the frequency) that the SPDC is triggered for the timestamps 1.1, 1.2 and 1.333~s. The communication network simulator is then rerun with the information that the SPDC triggers at these timestamps. It then computes new pdf's that we call \(pdf_1\). Fig.~\ref{fig:it1-800kbps} compares \(pdf_0\) and \(pdf_1\) for those three timestamps and for loads 8, 23, 29 and 39. Those loads are the loads located furthest away from the SPDC in each of the four zones defined in Fig.~\ref{fig:IEEE39}. Fig.~\ref{fig:it1-1.6Mbps} does the same comparison but for the case in which the bandwidth of the links has been set to 1.6~Mbps.

\begin{figure}
\centering
\begin{tikzpicture} [scale=0.7]
\begin{axis}[
    ybar,
    enlargelimits=0.2,
    legend style={at={(0.5,-0.15)},
      anchor=north,legend columns=-1},
    ylabel={Delays [ms]},
    symbolic x coords={Load 8, Load 23, Load 29, Load 39},
    xtick=data,
    ybar=2pt,
	bar width=8pt,
    ]
\addplot coordinates {(Load 8, 92.7957) (Load 23, 65.7581) (Load 29, 110.945) (Load 39, 128.375)};
\addplot coordinates {(Load 8, 92.7951) (Load 23, 65.7575) (Load 29, 110.945) (Load 39, 128.375)};
\addplot coordinates {(Load 8, 92.7953) (Load 23, 65.7578) (Load 29, 111.635) (Load 39, 128.375)};
\addplot coordinates {(Load 8, 99.5692) (Load 23, 72.5316) (Load 29, 118.475) (Load 39, 132.234)};
\legend{\(pdf_0\), \(pdf_1 (1.1)\), \(pdf_1 (1.233)\), \(pdf_1 (1.333)\)}
\end{axis}
\end{tikzpicture}
\caption{Comparison of the averages of the input and output pdf's at the first iteration of the self-consistent simulation. 800~kbps case}
\label{fig:it1-800kbps}
\end{figure}
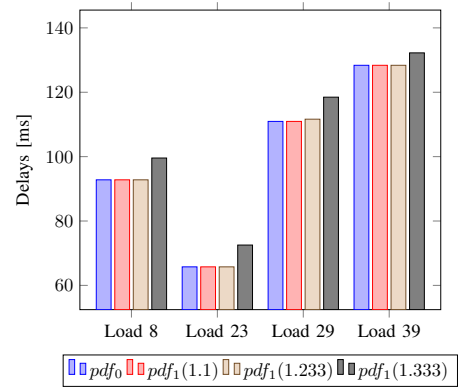

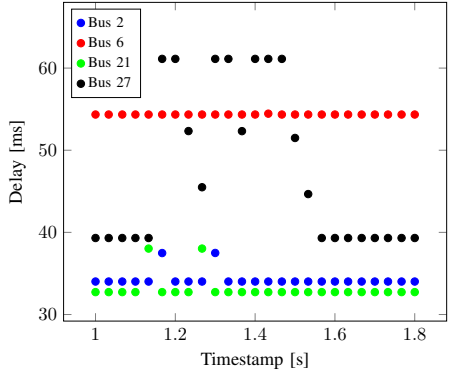
\begin{figure}
\centering
\begin{tikzpicture} [scale=0.7]
\pgfplotsset{width=\linewidth,
        legend style={font=\footnotesize}}
\begin{axis}[
    xlabel={Timestamp [s]},
    ylabel= {Delay [ms]},
    ymax=68,
    legend cell align=left,
    legend style={at={(0.02,0.98)},anchor=north west,
    },
   ]
    
  \addplot[only marks, blue] table [x=Timestamp,y=2] {Congestion.txt};
  \addplot[only marks, red] table [x=Timestamp,y=6] {Congestion.txt};
  \addplot[only marks, green] table [x=Timestamp,y=21] {Congestion.txt};
  \addplot[only marks] table [x=Timestamp,y=27] {Congestion.txt};
  
  \legend{Bus 2, Bus 6, Bus 21, Bus 27}

\end{axis}
\end{tikzpicture}
\caption{Delay between a measure being sent by bus \(i\) and being received by the SPDC. 800~kbps case}
\label{fig:congestion}
\end{figure}

As can be seen from the two figures, there is a non-negligible difference between \(pdf_0\) and \(pdf_1\) in the 800~kbps case (especially for the third control message), but not in the 1.6~Mbps case. The maximum difference (over all times and loads) between \(pdf_0\) and \(pdf_1\) is 7.5~ms in the first case, and 0.01~ms in the second. Fig.~\ref{fig:congestion} explains why these differences vary that much with the bandwidth. It shows the delays between a measure being sent by a bus and being received by the SPDC in the 800~kbps case. It shows that it is mostly the delay of measures coming from bus 27 (PDC 2) that is impacted by the additional traffic. This is because the link from bus 17 to bus 27 is saturated in normal operation. Indeed, monitoring traffic from 5 PMUs (at buses 14, 15, 16, 17 and 24) transits in this link. The total bandwidth used for this traffic is given by

\begin{equation}
    BW_{17\longrightarrow 27} = \text{\#PMU} \times f \times S = 600\text{ kbps}
\end{equation}

where \(f\) is the sampling frequency of the PMUs, and \(S\) the packet size in bits.

The monitoring traffic thus takes 75\% of the bandwidth of this link which causes additional traffic to have a large impact on the delay. For the other PDCs, the control traffic does not interfere as much with the monitoring traffic since they tend to go in opposite directions (and we use full-duplex links). In the 800~kbps case, the increased delay of measurements from PDC 2 (bus 27) makes this PDC a bottleneck and causes control messages to be sent later by the SPDC. In the 1.6~Mbps case, monitoring traffic only uses 37.5\% of the bandwidth of the link from bus 17 to bus 27. Additional traffic has thus a lower impact, and PDC 3 (bus 6) stays the bottleneck. The delay of measures coming from PDC 3 is (almost) independent of the additional traffic. The SPDC thus takes its decisions after an (almost) constant delay.

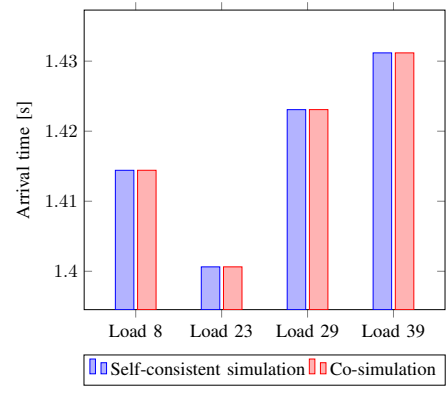
\begin{figure}
\centering
\begin{tikzpicture} [scale=0.7]
\begin{axis}[
    ybar,
    enlargelimits=0.2,
    legend style={at={(0.5,-0.15)},
      anchor=north,legend columns=-1},
    ylabel={Arrival time [s]},
    symbolic x coords={Load 8, Load 23, Load 29, Load 39},
    xtick=data,
    ybar=2pt,
    ]
\addplot coordinates {(Load 8, 1.4144157) (Load 23, 1.4006281) (Load 29, 1.4230751) (Load 39, 1.4311806)};
\addplot coordinates {(Load 8, 1.41442) (Load 23, 1.40063) (Load 29, 1.42308) (Load 39, 1.43118)};
\legend{Self-consistent simulation, Co-simulation}
\end{axis}
\end{tikzpicture}
\caption{Comparison of the arrival time of the third control packet between self-consistent and co-simulation. 1.6~Mbps case}
\label{fig:Compare1.6Mbps}
\end{figure}

We have shown that in the 1.6~Mbps case, communication network performance is not significantly affected by the additional traffic. In this case, our method converges in a single iteration. And Fig.~\ref{fig:Compare1.6Mbps} shows that our method gives coherent results compared to co-simulation. Indeed, it shows that the arrival times of the third load reduction message (i.e. that is sent when the frequency drops below the lowest frequency threshold) computed with the two approaches are almost identical. Similar results are obtained for the other load reduction messages. The maximum discrepancy is 0.02~ms. Note that to avoid synchronisation errors impacting the comparison, the arrival times in the co-simulation are derived from the communication network simulator. In the power system simulator, these events would be seen with up to 1~ms delay. This is not the case with the self-consistent approach.

\begin{figure}
\centering
\begin{tikzpicture} [scale=0.7]
\begin{axis}[
    ybar,
    enlargelimits=0.2,
    legend style={at={(0.5,-0.15)},
      anchor=north,legend columns=-1},
    ylabel={Delays [ms]},
    symbolic x coords={Load 8, Load 23, Load 29, Load 39},
    xtick=data,
    ybar=2pt,
	bar width=8pt,
    ]
\addplot coordinates {(Load 8, 92.5457) (Load 23, 65.5081) (Load 29, 110.895) (Load 39, 128.325)};
\addplot coordinates {(Load 8, 100.619) (Load 23, 72.8516) (Load 29, 118.845) (Load 39, 134.304)};
\addplot coordinates {(Load 8, 100.619) (Load 23, 72.8516) (Load 29, 118.845) (Load 39, 134.304)};
\legend{\(pdf_0\), \(pdf_1 (1.333)\), \(pdf_2 (1.333)\)}
\end{axis}
\end{tikzpicture}
\caption{Comparison of the averages of \(pdf_0\), \(pdf_1\) and \(pdf_2\) for the third load reduction message. 800~kbps case}
\label{fig:it2-800kbps}
\end{figure}
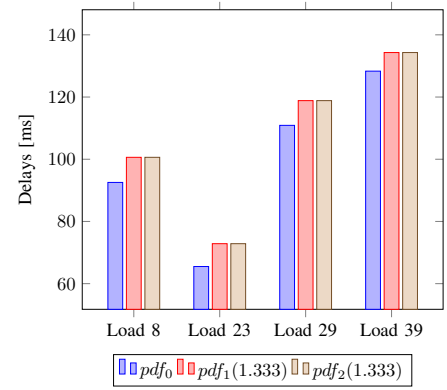

In the 800~kbps case, communication network performance is affected by the additional traffic, so we need to iterate. We thus use \(pdf_1\) as a model of the communication network in the power system simulator. However, we only computed a value of \(pdf_1\) for times 1.1, 1.233, and 1.333. So, if in the new power system simulation, a threshold is crossed at a different time, we use a linear interpolation to deduce a plausible value for the delay of a packet sent at that time. However, in our case, the threshold crossings occur at the same timestamps in the first and second iterations. \(pdf_1\) and \(pdf_2\) are thus identical (since they are computed with the same additional traffic), as shown in Fig~\ref{fig:it2-800kbps}. Our method thus converges after this second iteration. Again, we can compare the arrival time of control packets computed with self-consistent and co-simulation, and the maximum discrepancy is 0.01~ms.

We now compare the computational burden of self-consistent and co-simulation for the 800~kbps case. Table~\ref{tab:computationTime} shows the computation time needed to perform a 5s simulation with both methods\footnote{Note that both Dynawo and ns-3 are single-threaded programs.}. Computations were performed using an AMD 4500U processor. This table compares the two approaches for different ``time precisions''. In co-simulation, time precision refers to the minimum time between synchronisation steps. In self-consistent simulation, it refers to the threshold used in the stopping criterion. These two concepts of errors are quite different. Indeed, in co-simulation errors are additional ``delays'' introduced at the interface between the simulators. Co-simulation thus tends to give conservative results. On the other hand, errors in self-consistent simulation are caused by the difference between \(pdf_i\) and \(pdf_{i+1}\). They can thus be conservative or optimistic. We thus recommend using a (slightly) lower time precision when using self-consistent simulation. However, as shown by the table, self-consistent simulation is faster to perform than co-simulation, even if using a more accurate time precision. Self-consistent simulation is particularly effective at high time precision since it does not have to perform frequent synchronisations. Indeed, it shows a speed-up of a factor 15 for a time precision of 1~ms.

\begin{table} 
\centering
\caption{Comparison of computation times using self-consistent and co-simulation. The simulated time is 5~s. 800~kbps case}
\begin{tabular}{@{}lll@{}}
\toprule
Time precision & Self-consistent simulation & Co-simulation \\ \midrule
10 ms     & 4 s                        & 21 s          \\
1 ms      & 7 s                        & 110 s         \\ \bottomrule
\end{tabular}
\label{tab:computationTime}
\end{table}


\section{Conclusions}
\label{sec:Conclusion}

In this work, we proposed self-consistent simulation as an alternative to co-simulation for simulating the interactions of transmission systems with their ICT infrastructures. We then applied this approach to the IEEE 39-bus test system equipped with an all-PMU state estimator. The test system also included a simple control algorithm based on this state estimator. Using this test case, we showed that our approach gives coherent results compared to co-simulation. We also showed that our approach can be up to one order of magnitude faster than co-simulation especially for cases where there are numerous interactions between the power system and ICT domain. This speed-up should be welcomed in applications where lots of simulations are performed such as in sensitivity analyses or probabilistic security assessments.


Beyond speed, our method has several advantages compared to co-simulation. First, in the cases where co-simulation can be completely avoided, it only requires the power system simulator to be able to simulate (pseudo) random delays (with known pdf's). It should thus be easier to implement than co-simulation (which requires the simulators to frequently pause to exchange information), especially when using closed-source simulators. Another advantage of the proposed methodology is that the pdf's of the delays do not necessarily have to be computed using a communication network simulator, but they can also be obtained from historical data and the large literature of traffic engineering models. Finally, if the pdf's are modified (e.g. to perform a sensitivity analysis), it is possible to keep previously simulated samples of those pdf's which reduces the number of new simulations to be performed.

In future work, we will assess how our approach performs when the communication network is operating in more degraded states such as during cascading outages, and considering variable reporting rates. Also, we will consider cases with uncertainties and/or errors models such that it is required to use pdf's instead of Dirac distributions.


\bibliographystyle{IEEEtran}
\bibliography{IEEEabrv,bib}

\begin{thebibliography}{10}
\providecommand{\url}[1]{#1}
\csname url@samestyle\endcsname
\providecommand{\newblock}{\relax}
\providecommand{\bibinfo}[2]{#2}
\providecommand{\BIBentrySTDinterwordspacing}{\spaceskip=0pt\relax}
\providecommand{\BIBentryALTinterwordstretchfactor}{4}
\providecommand{\BIBentryALTinterwordspacing}{\spaceskip=\fontdimen2\font plus
\BIBentryALTinterwordstretchfactor\fontdimen3\font minus
  \fontdimen4\font\relax}
\providecommand{\BIBforeignlanguage}[2]{{%
\expandafter\ifx\csname l@#1\endcsname\relax
\typeout{** WARNING: IEEEtran.bst: No hyphenation pattern has been}%
\typeout{** loaded for the language `#1'. Using the pattern for}%
\typeout{** the default language instead.}%
\else
\language=\csname l@#1\endcsname
\fi
#2}}
\providecommand{\BIBdecl}{\relax}
\BIBdecl

\bibitem{CoSimChallenges}
{IEEE Task Force on Interfacing Techniques for Simulation Tools}, ``Interfacing
  power system and {ICT} simulators: Challenges, state-of-the-art and case
  studies,'' \emph{IEEE Transactions on Smart Grid}, vol.~9, no.~1, pp. 14--24,
  2018.

\bibitem{VogtCosimReview}
M.~Vogt, F.~Marten, and M.~Braun, ``A survey and statistical analysis of smart
  grid co-simulations,'' \emph{Applied Energy}, vol. 222, pp. 67--78, 2018.

\bibitem{GridAttackSim}
\BIBentryALTinterwordspacing
T.~D. Le, A.~Anwar, S.~W. Loke, R.~Beuran, and Y.~Tan, ``{GridAttackSim}: A
  cyber attack simulation framework for smart grids,'' \emph{Electronics},
  vol.~9, no.~8, 2020. [Online]. Available:
  \url{https://www.mdpi.com/2079-9292/9/8/1218}
\BIBentrySTDinterwordspacing

\bibitem{ToolsForCosim}
\BIBentryALTinterwordspacing
R.~Czekster, ``Tools for modelling and simulating the smart grid,'' \emph{arXiv
  Preprints}, 2020. [Online]. Available:
  \url{https://arxiv.org/pdf/2011.07968v3.pdf}
\BIBentrySTDinterwordspacing

\bibitem{GECOcomputationTime}
H.~Lin, S.~S. Veda, S.~S. Shukla, L.~Mili, and J.~Thorp, ``{GECO}: Global
  event-driven co-simulation framework for interconnected power system and
  communication network,'' \emph{IEEE Transactions on Smart Grid}, vol.~3,
  no.~3, pp. 1444--1456, 2012.

\bibitem{TrafficVolumeBook}
K.~C. Budka, J.~G. Deshpande, and M.~Thottan, \emph{An Overview of Smart Grid
  Network Design Process}.\hskip 1em plus 0.5em minus 0.4em\relax London:
  Springer London, 2014, pp. 169--207.

\bibitem{GECOtestcase}
H.~Lin, Y.~Deng, S.~Shukla, J.~Thorp, and L.~Mili, ``Cyber security impacts on
  all-{PMU} state estimator – {A} case study on co-simulation platform
  {GECO},'' in \emph{2012 IEEE Third International Conference on Smart Grid
  Communications (SmartGridComm)}, 2012.

\bibitem{IEEE39}
T.~{Athay}, R.~{Podmore}, and S.~{Virmani}, ``A practical method for the direct
  analysis of transient stability,'' \emph{IEEE Transactions on Power Apparatus
  and Systems}, vol. PAS-98, no.~2, pp. 573--584, 1979.

\bibitem{IEEE39transient}
P.~Demetriou, M.~Asprou, J.~Quiros-Tortos, and E.~Kyriakides, ``Dynamic {IEEE}
  test systems for transient analysis,'' \emph{IEEE Systems Journal}, vol.~11,
  no.~4, pp. 2108--2117, 2017.

\bibitem{Monolithic}
D.~Jafarigiv, K.~Sheshyekani, H.~Karimi, and J.~Mahseredjian, ``A scalable
  {FMI}-compatible cosimulation platform for synchrophasor network studies,''
  \emph{IEEE Transactions on Industrial Informatics}, vol.~17, no.~1, pp.
  270--279, 2021.

\bibitem{LineLength}
C.~K. Das, T.~S. Mahmoud, O.~Bass, S.~Muyeen, G.~Kothapalli, A.~Baniasadi, and
  N.~Mousavi, ``Optimal sizing of a utility-scale energy storage system in
  transmission networks to improve frequency response,'' \emph{Journal of
  Energy Storage}, vol.~29, p. 101315, 2020.

\bibitem{StandardC37-118-2}
{IEEE}, ``{IEEE} standard for synchrophasor data transfer for power systems,''
  \emph{IEEE Std C37.118.2-2011 (Revision of IEEE Std C37.118-2005)}, pp.
  1--53, 2011.

\bibitem{dynawo}
A.~Guironnet, M.~Saugier, S.~Petitrenaud, F.~Xavier, and P.~Panciatici,
  ``Towards an open-source solution using {M}odelica for time-domain simulation
  of power systems,'' in \emph{2018 {IEEE} {PES} Innovative Smart Grid
  Technologies Conference Europe ({ISGT}-Europe)}, 2018.

\bibitem{ns3}
T.~R. Henderson, M.~Lacage, G.~F. Riley, C.~Dowell, and J.~Kopena, ``Network
  simulations with the ns-3 simulator,'' \emph{SIGCOMM demonstration}, vol.~14,
  no.~14, p. 527, 2008.

\bibitem{helics}
B.~Palmintier, D.~Krishnamurthy, P.~Top, S.~Smith, J.~Daily, and J.~Fuller,
  ``Design of the {HELICS} high-performance
  transmission-distribution-communication-market co-simulation framework,'' in
  \emph{2017 Workshop on Modeling and Simulation of Cyber-Physical Energy
  Systems (MSCPES)}, 2017.

\end{thebibliography}

\end{document}